\documentclass[12pt, a4paper]{article}

\begin{document}

\title{QUANTUM COHERENCE, CORRELATED NOISE AND PARRONDO GAMES}

\author{CHIU FAN LEE and NEIL F. JOHNSON \\
Center for Quantum Computation and Physics Department\\
Clarendon Laboratory, Parks Road\\
Oxford OX1 3PU, U.K.\\
\\
FERNEY RODRIGUEZ and LUIS QUIROGA \\
Departamento de Fisica, Universidad de Los Andes\\
Bogota, A.A. 4976, Colombia}

\maketitle

\begin{abstract} We discuss the effect of correlated noise on the
robustness of quantum coherent phenomena. First we consider a simple,
toy model to illustrate the effect of such correlations on the
decoherence process. Then we show how decoherence rates can be suppressed
using a Parrondo-like effect. Finally, we report the results of
many-body calculations in which an experimentally-measurable quantum
coherence phenomenon is significantly enhanced by non-Markovian dynamics
arising from the noise source.
\end{abstract}

\section{Introduction}  Decoherence is a uniquely quantum phenomenon which
results in the decay of the off-diagonal elements of the system's density
matrix. This implies that quantum superposition states decay into a
probabilistic mixture. Decoherence, and in particular the control of
decoherence, is crucial to the success of quantum information processing
and quantum computation
\cite{Nielsen}. Decoherence can be considered as arising due to the
presence of some kind of `noise'. Regardless of its origin, such `noise'
is typically assumed to be a statistically independent stochastic process, i.e. lacking
any temporal correlations. The  off-diagonal elements of the density matrix then tend
to decay in time as a simple exponential, with the decay constant
denoting a `relaxation time'.  However, recent work has shown that such
approximations are highly questionable - or even plainly wrong - in the
ultrafast optical regime being explored for nanostructure-based quantum
information processing~\cite{Ferney,Luis}.

In this Letter we investigate the effect of correlated noise on quantum
coherent phenomena. First  we consider a simple, toy model of a two-level
system in order to illustrate the effect that such correlations can have
on decoherence. Then we show how decoherence rates can be suppressed
via a Parrondo-like effect~\cite{PHA00}. Finally, we report the results of
detailed many-body calculations in which an experimentally-measurable
quantum coherent phenomenon is significantly enhanced by non-Markovian
dynamics arising from the noise source. Our results show that, in addition
to yielding  non-trivial dynamics, correlated noise can play an important
role in determining the dynamics of the decoherence process.

\section{Toy Model of Decoherence}   We start by considering decoherence
of a two-level system under the physical assumptions that the channel is
isolated (i.e. no entanglement between the system and the environment) and
non-dissipative (i.e. only phase damping). We further assume discrete
time-evolution for simplicity. Our model is a generalization of Nielsen
and Chuang's  treatment of a single phase-kick~\cite{Nielsen}.

Suppose we have a qubit
$|\Psi\rangle_0 = a_0|0\rangle + b_0|1\rangle$ at timestep $t=0$. A
rotation
$R_z(\theta_1)$ is applied at timestep $t=1$, to model a phase-kick.  The
random angle
$\theta_1$  is drawn from a probability distribution $P_1(\theta_1)$. The
output state is given by the density matrix obtained from averaging over
$\theta_1$:
\begin{equation}
\rho_1 = \int_{-\pi}^\pi R_z(\theta_1)
\rho_0 R_z^\dag(\theta_1) P_1(\theta_1) d \theta_1 \ \ .
\end{equation}  The density matrix after $n$ timesteps is given by:
$$\rho_n =
\int_{-\pi}^\pi
\dots\int_{-\pi}^\pi
\int_{-\pi}^\pi
 R_z(\theta_n) \dots R_z(\theta_2) R_z(\theta_1) \rho_0\ \dots\ \ \ \ \ \
\ \ \ \ \ $$
\begin{equation}
\ \ \ \ \\ \ \ \ \ \ \ \ \ \ \ \  R_z^\dag (\theta_1) R_z^\dag (\theta_2)
\dots R_z^\dag (\theta_n)  P(\theta_n, \theta_{n-1}, \dots,\theta_2,
\theta_1) d\theta_1 d
\theta_2 \dots d\theta_n \ \
\end{equation} where Bayes' rule yields
\begin{equation} P(\theta_n, \theta_{n-1}, \dots,\theta_2, \theta_1) =
\prod_{i=1}^{n} P(\theta_i |\theta_{i-1}, \theta_{i-2}, \dots, \theta_2,
\theta_1) \ \ .
\end{equation}  $R_z(\theta_n) R_z(\theta_{n-1})\dots R_z(\theta_2)
R_z(\theta_1)$ can be rewritten exactly as a single net rotation
$R_z(\Theta)$ where $\Theta = \sum_{i=1}^n \theta_i$. In general the
stochastic process producing the kicks may have temporal correlations,
and may not be stationary - hence the kicks may be dependent and/or have
a time-dependent probability distribution (e.g. if the noise-source were
changing in time).

We start by considering {\em independent, identically distributed} ({\em
i.i.d.}) kicks. Hence
$P(\theta_i|\theta_{i-1},\dots)=P_i(\theta_i)$ with
$P_i(\dots)=P(\dots)$ for all
$i$. Thus
\begin{equation} P(\theta_n, \theta_{n-1}, \dots,\theta_2, \theta_1) =
\prod_{i=1}^{n} P(\theta_i)
\end{equation} If we also assume that $P(\theta)$ is Gaussian, then
$P(\theta) = \frac{1}{{\sqrt{4\pi\lambda}}}
e^{-{\frac{\theta^2}{4\lambda}}}$ where the mean is 0 and the variance is
$2\lambda$. With the assumption that the Gaussian distributions have narrow
peaks, we can allow the lower (upper) limits of the integrals to go to
minus infinity (plus infinity). 
This yields $\rho_n$ as a matrix with off-diagonal elements
$a_0b_0^* e^{-n\lambda}$. This exponential {linear} dependence on $n$ is
basically just the effect of $n$ independent Gaussian integrals.
Equivalently we can think of the net rotation $\Theta =\sum_{i=1}^n
\theta_i$ as having a variance $2\lambda_n$ given by the sum of the
variances
$2\lambda_n=2\lambda n$. For non-Gaussian $P(\theta)$, it can be shown
that an exponential decay also arises.

We now consider {\em dependent}, but identically distributed ({\em
d.i.d.}) kicks. As an illustration, we consider the simple case of the
angles being so strongly correlated that $\theta_2=\theta_1$,
$\theta_3=\theta_2$, etc. This is equivalent to saying that all the
conditional probabilities
$P(\theta_i|\theta_{i-1},
\theta_{i-2},
\dots, \theta_2, \theta_1)=\delta(\theta_i-\theta_1)$. Assuming that $\theta_1$ has a
Gaussian distribution, then
$P(\theta) = \frac{1}{{\sqrt{4\pi\lambda}}}
e^{-{\frac{\theta^2}{4\lambda}}}$.  Now $\rho_n$ has off-diagonal elements
$a_0b_0^* e^{-n^2\lambda}$. This exponential {\em quadratic} dependence on
$n$ is basically just the effect of $n$ dependent integrals. Equivalently
we can think of the net rotation
$\Theta =\sum_{i=1}^n \theta_i=n\theta_1$ as having a variance
$2\lambda_n$ given by
$2\lambda_n=2\lambda n^2$. Hence the noise correlations have
significantly affected the decoherence process. 

\section{Decoherence Control using Parrondo Effect} 
We now proceed to
investigate the counter-intuitive effect whereby noise correlations
might be exploited in an active way, in order to {\em reduce} decoherence.
The Parrondo effect
\cite{PHA00} is a remarkable result whereby two losing `games', when
combined, become winning. Pioneered by J.M.R. Parrondo, and by D. Abbott
and collaborators, the most striking feature is arguably the fact that
this combination can be {\em random}, i.e. random switching between two
losing games A and B can produce a winning game C. Several realizations
of the Parrondo effect have recently been suggested in the quantum
regime: in particular, quantum games \cite{Adrian}, quantum lattice gases
\cite{Meyer} and quantum algorithms \cite{ChiuFan}. But couldn't the same
idea be applied to quantum decoherence? In particular, could two
decoherence sources (two `private baths') be combined to produce a single
decoherence source (a `public bath') with a longer coherence time?

For independent kicks (either identically or non-identically distributed)
we have been able to show that it is {\em not} possible to produce such a Parrondo
effect~\cite{ChiuFan}.  Remarkably, however, we {\em can} produce a Parrondo-like effect
if we allow for correlated kicks, i.e. correlated noise. We
now illustrate this, using a specific example motivated by the classical
vector-rotating game of Ref.~
\cite{ChiuFan}. Consider two probability distributions
$P_A, P_B$ corresponding to two `private baths' for phase-damping kicks.
These distributions are such that the kick rotation angle $\theta_2$ is correlated
to the previous rotation angle ($\theta_1$) in the following manner:
\begin{eqnarray}
P_A(\theta_2|\theta_1) &=&
\left\{ \begin{array}{ll}
\frac{1}{3}[ \delta(\theta_2) +
\delta(\theta_2+\frac{\pi}{2}) + \delta(\theta_2-\frac{\pi}{2})]
 & , \ \theta_1 \in \{-\frac{\pi}{2}, 0, \frac{\pi}{2} \} \\
\delta(\theta_2) & ,\ {\rm otherwise}
\end{array} \right. \\
 P_B(\theta_2|\theta_1) &=&
\left\{ \begin{array}{ll}
\frac{1}{3}[ \delta(\theta_2-\epsilon)
 +\delta(\theta_2+\frac{3 \pi}{4}) + \delta(
 \theta_2-\frac{\pi}{4} ) ]
 & , \ \theta_1 \in \{-\frac{3 \pi}{4}, \epsilon, \frac{\pi}{4} \} \\
\delta(\theta_2-\epsilon) & ,\ {\rm otherwise}
\end{array} \right.
\end{eqnarray}
with similar conditions holding for all subsequent pairs $\theta_i$ and
$\theta_{i-1}$. We note that the specific choice of angles is designed to make
analytical calculations  possible, and may be generalized. The parameter
$\epsilon$ is supposed to be small---its presence serves to `memorize' which
probability distribution was selected in the previous step. If $P_A$ represents the only
noise-source applied to the system, and assuming the initial angle of rotation is $0$
(i.e.
$\theta_1=0$) then we have
\begin{equation}
P_A(\theta_n, \ldots, \theta_1)= \prod_{i=2}^{n} P_A(\theta_{i}|
\theta_{i-1})= (\frac{1}{3})^{n-1}
\end{equation} 
since the angles always lie in the set  $\{-\pi/2, 0,\pi/2\}$.
If the initial density matrix is
\begin{equation}
\rho_1:=
\left(
\begin{array}{cc}
a & b \\ b^\ast & d 
\end{array} \right),
\end{equation}
then the density matrix after $n$ time steps is:
\begin{equation}
\rho_n:=
\left(
\begin{array}{cc}
a & b\gamma^n e^{-in\phi} \\ b^\ast \gamma^n e^{in\phi} & d 
\end{array} \right),
\end{equation}
where 
\begin{equation}
\gamma e^{\pm i\phi}:= \frac{1}{3}
\end{equation}
assuming that the system is under the influence of $P_A$ alone. 
Similar arguments hold if $P_B$ is the only noise-source applied to the system and
if we assume $\theta_1 =\epsilon$. In this case,
\begin{equation}
\gamma e^{\pm i\phi}:= \frac{1}{3}e^{\pm i \epsilon}.
\end{equation}

Combining the two noise-sources (i.e. probability distributions) at random gives
\begin{equation} P(\theta_2|\theta_1) =
\left\{ \begin{array}{ll}
\frac{1}{2}\delta(\theta_2-\epsilon)
 + \frac{1}{6} [\delta(\theta_2-0) +\delta(\theta_2+\frac{\pi}{2}) + 
 \delta(\theta_2-\frac{\pi}{2}
)] 
 & , \ \theta_1 \in \{-\frac{\pi}{2}, 0,\frac{\pi}{2} \} \\
\\
\frac{1}{2}\delta(\theta_2-0)
 + \frac{1}{6} [\delta(\theta_2-\epsilon)
 +\delta(\theta_2+\frac{3 \pi}{4}) + \delta(\theta_2-\frac{\pi}{4}) ]
 & , \ \theta_1 \in \{-\frac{3 \pi}{4}, \epsilon, \frac{\pi}{4} \}\\
\\
\frac{1}{2}[\delta(\theta_2) +\delta(\theta_2-\epsilon)]
&, \ {\rm otherwise}. \\
\end{array} \right.
\end{equation} The density matrix $\rho_n$ is given by
\begin{equation}
\int R_z(\theta_1) \cdots \int R_z(\theta_n) \rho_0 R_z^\dag (\theta_n)
P(\theta_n|\theta_{n-1}) d \theta_n
\cdots R_z^\dag (\theta_1) P(\theta_1) d \theta_1
\end{equation}
 where $P$ only has one-step correlations. We define the following
functions recursively:
\begin{eqnarray} f_1(\theta) &:=& \int e^{i \phi}P(\phi| \theta) d \phi
\\ f_{k+1}(\theta) &:=&
\int e^{i\phi}f_k(\phi) P(\phi| \theta) d \phi
\end{eqnarray} for $1\leq k\leq n$. Assuming the initial angle is 0,
we have
\begin{equation}
\rho_n:=
\left(
\begin{array}{cc}
a & b [f_n(0)]^\ast \\ b^\ast f_n(0) & d 
\end{array} \right).
\end{equation}
For the combined probability distribution $P$ above,
the angles of rotation can only take on seven values,
$\{ -\pi/3 , -\pi/2, 0,
 \epsilon, \pi/3, \pi/2, \pi \}$. We can calculate the $f_k$'s as follows:
\begin{eqnarray} f_1 &:=&
\left\{ \begin{array}{ll}
\{-\pi/2, 0, \pi/2 \} & \mapsto \ e^{i \epsilon}/2 + 1/6 \\
\{ -3 \pi/4 , \epsilon, \pi/4 \} & \mapsto \ 1/2 + e^{i \epsilon}/6
\end{array} \right. \\
 f_{k+1} &:=&
\left\{ \begin{array}{ll}
\{-\pi/2, 0, \pi/2 \} & \mapsto \
\frac{e^{i \epsilon}}{2} f_k(\epsilon) +
\frac{1}{6} f_k(0) \\
\{ -3 \pi/4 , \epsilon, \pi/4 \} & \mapsto \ 
\frac{1}{2} f_k(0) + \frac{e^{i \epsilon}}{6} f_k( \epsilon).
\end{array} \right.
\end{eqnarray} Letting $\epsilon$ go to zero and writing $e^{i \epsilon}$ as
$1 + {\cal O}(\epsilon)$, we see that 
$f_1(0)=f_1(\epsilon)= 2/3 +{\cal O}(\epsilon)$,
$f_2(0)= \frac{1}{2}f_1(\epsilon) +  \frac{1}{6}f_1(0) =
(2/3)(2/3) + {\cal O}(\epsilon)$, and so on. Indeed, $f_k(0)= (2/3)^{k-1}+
{\cal O}(\epsilon)$ for all $k$.
Hence the decay factor $\gamma$ here is $2/3
+ {\cal O}(\epsilon)$, which is an {\em improvement} over $1/3$.  This
reflects the Parrondo effect. Hence our toy model has established that, in
principle, it might be possible to change decoherence properties in a favourable way
by {\em adding} more noise to a quantum system. Our toy model has also demonstrated
that temporal correlations can have a significant
effect on decoherence properties and hence must be accounted
for properly in any realistic calculations.

\section{Enhancement of Quantum Coherent Phenomena} The
possibility of performing quantum information processing in nanostructure
systems, such as semiconductor quantum dots (QD), is of great interest
from the perspectives of both fundamental science and future emerging
technologies~\cite{Luis}. Significant advances have been made recently in the
fabrication of such nanostructures. 
Apart from quantum dots, there are
a wide range of inorganic and organic structures which also qualify as `nanostructures',
including microbiological molecular structures such as the photosynthetic complexes in
purple bacteria.
However the great challenge facing any
such information processing in the quantum regime lies in avoiding, controlling or
overcoming the effects of decoherence. Typical decoherence times are of the order of
picoseconds, and hence impose severe constraints on the timscales within which quantum
logic gates need to be perfomed. For this reason, it is now
widely believed that excitons generated optically in such
nanostructures, could serve as useful qubits. Their manipulation (i.e. quantum logic
gates) could then be achieved by ultrafast femtosecond laser pulses. Remarkably, such
manipulation has already been demonstrated experimentally for single quantum dot
nanostructures~\cite{bonadeo}. 

Given the potential importance of correlated noise on decoherence effects 
demonstrated in this Letter, we have investigated the effect of non-Markovian dynamics
(i.e. correlated noise) in such nanostructure systems. In particular, we performed
large-scale, many-body calculations of the ultrafast second-order coherence function of
the emitted light from the optically-generated exciton in a single nanostructure (QD).
Non-Markovian effects are included for both exciton-photon and exciton-phonon
couplings. We find that a strong photon antibunching effect (a purely quantum
phenomenon) arises in the resonance fluorescence response at very short times, if and
only if the initial exciton state comprises a quantum superposition
\cite{Ferney}. More importantly for the present study, we find that
correlation effects significantly enhance the antibunching signal,
hence demonstrating explicitly that temporal correlations {\em cannot} be neglected
a priori in such ultrafast regimes. The typical Markov approach, via
Master equations with time-independent damping coefficients corresponding
to an exponential decay in decoherence (c.f. earlier discussion for {\em
i.i.d.} kicks), {\em cannot} account for the evolution of an open system
on very short time scales. In short, Markov approximations (which are
valid on long time-scales) overestimate the decay effects at short times. Hence great
care must be exercised when treating temporal correlation effects on such short
time-scales.

\section{Conclusion} We have discussed the effects of correlated noise
on quantum coherence, and have shown the possibility of decoherence
control through a Parrondo-like effect. We have also reported the crucial
role that  non-Markovian damping effects (temporal correlations)
can play in nanostructure-based quantum information processing.

\section*{Acknowledgements}  CFL thanks NSERC (Canada), ORS (U.K.) and
Clarendon Fund (Oxford) for financial support. FR and LQ acknowledge
support from COLCIENCIAS (Colombia) projects No.1204-05-10326,
1204-05-1148 and Banco de la Rep\'ublica (Colombia). NFJ thanks EPSRC for
a Travel Grant.


\end{document}